\documentclass{phep}       

\journal{PHEP}
\vol{xx}

\received{xx January 2025}
\published{xx March 2025}

\usepackage{amsmath}
\usepackage{amssymb}
\usepackage[hidelinks]{hyperref}
\usepackage[left]{lineno}

\def\be{\begin{equation}}
\def\ee{\end{equation}}
\def\bea{\begin{eqnarray}}
\def\eea{\end{eqnarray}}

\renewcommand{\thefootnote}{\fnsymbol{footnote}}

\begin{document}

\title{The Light Dark Matter eXperiment}

\author{Tamas Almos Vami\auno{1}\footnotemark\ for the LDMX Collaboration}
\address{$^1$University of California Santa Barbara, Santa Barbara, CA, USA}

\begin{abstract}
Searching for dark matter (DM) at colliders is one of the biggest challenges in high-energy physics today. Significant efforts have been made to detect DM within the mass range of 1-10,000~GeV at the Large Hadron Collider and other experiments. However, the lower mass range of 0.001-1~GeV remains largely unexplored, despite strong theoretical motivation from thermal dark matter models in that mass range. The Light Dark Matter eXperiment (LDMX) is a proposed fixed-target experiment at SLAC’s LCLS-II 8~GeV electron beamline, specifically designed for the direct production of sub-GeV dark matter. The experiment operates on the principle of detecting missing momentum and missing energy signatures. In this talk, we will present the experimental design of LDMX detector and discuss strategies for detecting dark matter. The talk will detail traditional discriminants-based methods using the electromagnetic and hadronic calorimeters as a veto for Standard Model processes. Additionally, the application of advanced machine learning techniques, such as boosted decision trees and graph neural networks, for distinguishing signal from background will be discussed.
\end{abstract}

\maketitle


\renewcommand{\thefootnote}{\fnsymbol{footnote}}
\footnotetext{Corresponding author: Tamas.Almos.Vami@cern.ch}

\section{Light dark matter production}

Characterizing dark matter (DM) is one of the biggest challenge of modern high-energy physics. Extensive searches at the Large Hadron Collider and other experiments have focused on DM particles with masses between 1 and 10,000~GeV~\cite{cms-summary,atlas-summary,NA64:2019imj}. In contrast, the lighter mass range of 0.001-1~GeV is still largely unexplored, despite strong theoretical motivation from thermal DM scenarios. In these models, DM is initially assumed to be in a thermal equilibrium with the Standard Model (SM) particles in the early universe. As the universe expands and cools, this equilibrium is broken through a freeze-out process, which determines the present-day DM abundance~\cite{CatenaGray2023}.

Given the constraints of the current DM abundance, we can calculate a production amplitude for an accelerator-based production, via dark photon (A') emission in electron–target interactions, analogous to the bremsstrahlung process. This calculation leads to the result that the dark bremsstrahlung is about 15 orders of magnitude more rare than the SM bremsstrahlung process. Consequently, by measuring on the order of $10^{14}$–$10^{16}$ electron–target interactions, one can place definitive constraints on thermal DM scenarios.

Starting in 2027 the SLAC linear accelerator will provide an 8~GeV electron beam for LCLS-II operations, and the End Station A (ESA) offers an ideal environment for fixed-target experiments~\cite{markiewicz2022slaclinacesalesa}. By diverting ultra–low-charge bunches ($\mu = 1$) to ESA, using the Sector 30 Transfer Line kicker, we can perform dedicated fixed-target collisions in the ESA experimental area. This setup enables the construction of a compact detector to reconstruct the final-state particles, known as the Light Dark Matter eXperiment (LDMX). The detailed design report for LDMX is available in Ref.~\cite{LDMX:2025rwc}.

\section{How to measure DM?}

In the dark bremsstrahlung process, the electron gets a transverse kick,
and the produced dark photon (and DM) is invisible. To characterize the DM, we rely on the missing-momentum and missing-energy measurements. We need to be able to reject all events where an SM photon is produced and potentially converted into other SM particles. 

\begin{figure}[!htb]
    \includegraphics[width=0.5\textwidth]{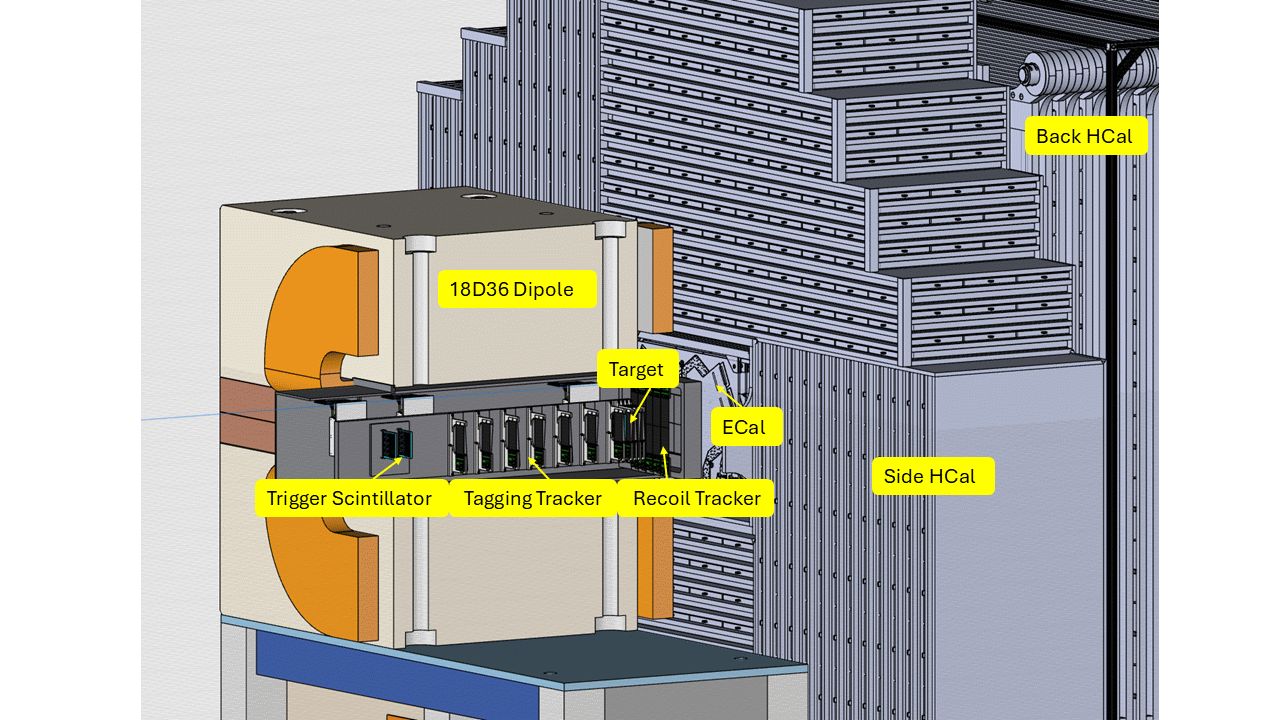}
    \caption{The LDMX detector showing the trigger scintillators, the tagging tracker, the target inside the spectrometer dipole, the recoil tracker, the ECal, and the side and back HCal.}
    \label{fig:LDMX_overview_cutaway}
\end{figure}

In order to achieve this, we need a fast, high momentum resolution tracker; a fast, high granularity, radiation tolerant electromagnetic
calorimeter; and a hermetic hadronic calorimeter. All these technologies are available, so we take them and customize them for the LDMX needs from other experiments. The schematic design of the LDMX detector is displayed in Fig.~\ref{fig:LDMX_overview_cutaway}.

The magnet and tracker design are adapted from the HPS experiment~\cite{hps_proposal_2014}. The magnet provides a 1.5~T magnetic field for the  momentum measurement. We use 7  double layers of the silicon strip tagger tracker to measure that the incoming electron's energy is compatible with the 8~GeV beam and to remove off-energy beam electrons. Then an $0.1X_0$ tungsten target is used to produce the dark photons. This is followed by 6 double layers forming the silicon strip recoil tracker. The recoil tracker is responsible to measure the $p_T$ of the electron after it receives its transverse kick from the (dark) photon production.

The electromagnetic calorimeter (ECal) is the small version of the CMS Phase-II upgrade's High Granularity Calorimeter~\cite{Milella:2023tkm,CMSHGCAL:2020dnm}. It is a sampling calorimeter with tungsten absorber and silicon sensors. We use 17 double layers with varying absorber thickness. The radius of the cell is about 5~mm, which allows for tracking minimum ionizing particles (MIPs) produced in the ECal.

The hadronic calorimeter (HCal) is based on Mu2e’s Cosmic Ray Veto~\cite{Mu2e:TDR}. It uses plastic scintillators and steel absorbers together with a SiPM readout, similar to the readout in CMS~\cite{HGCROC}. The HCal catches the leftover hadrons that were not found by the ECal, which are mostly neutrons.

\section{What are the backgrounds?}

\begin{figure}[!h]
    \includegraphics[width=0.42\textwidth]{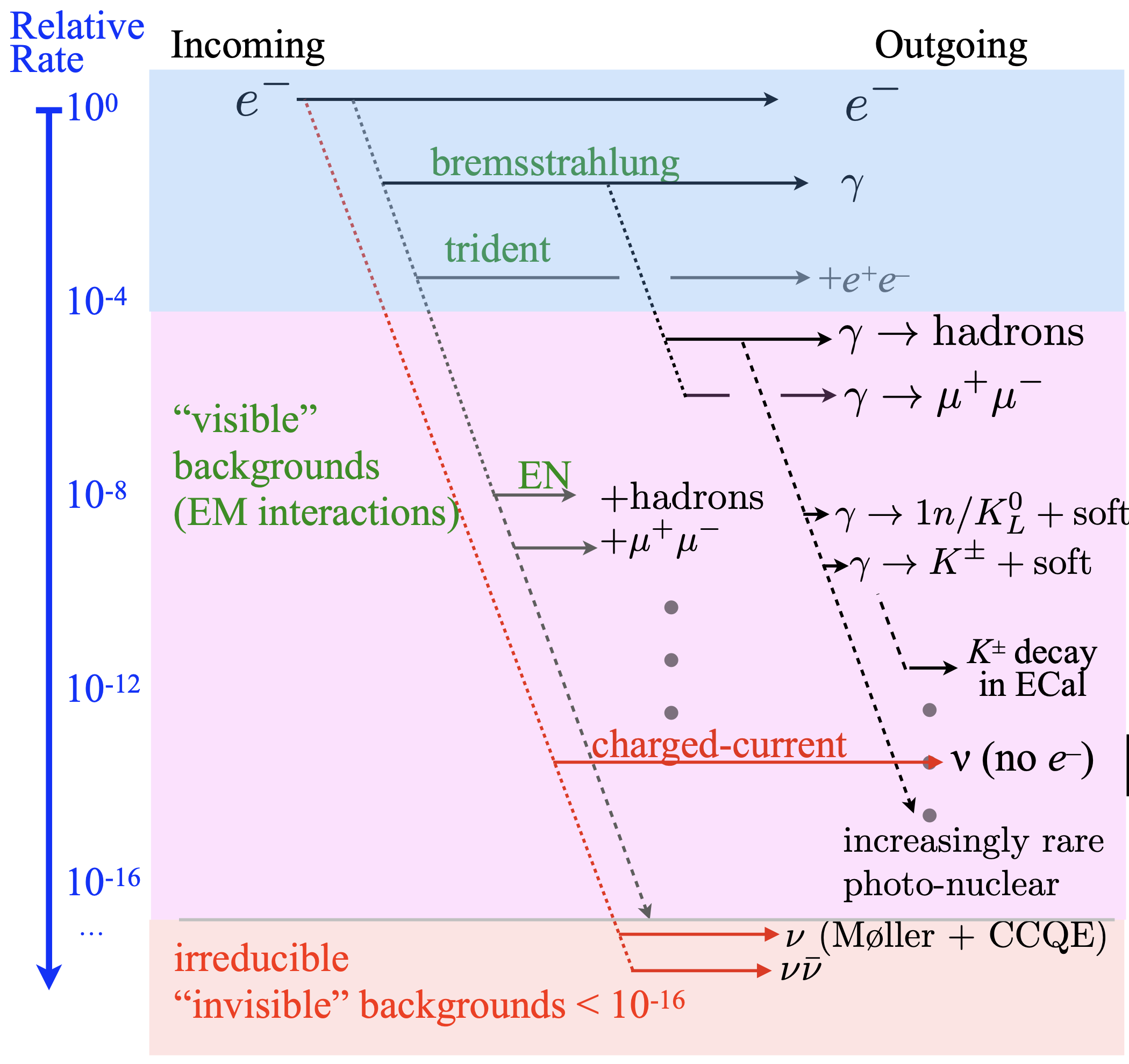}
    \caption{Relative rates of different SM processes that are the backgrounds in the experiment .}
    \label{fig:backgroundStaircase}
\end{figure}

We have to be able to reject $10^{14}$ background events. This is the equivalent scale of finding each single cell in the whole human body! Figure~\ref{fig:backgroundStaircase} shows the relative rate of different SM processes with respect to a simple non-interacting electron.

The most abundant process is the bremsstrahlung process, which we can easily reject by measuring the energy of the electron and photon in the ECal. When the photon converts into an electron positron pair at the target we can rely on the recoil tracker's measurement to reject events with more than 1 track. The most complicated backgrounds are the cases when a photonuclear reaction creates mesons and hadrons in the ECal. For these, we have to understand the showers in the ECal in great details,
and rely also on HCal to reject these. For the kaons produced in the ECal, we rely on tracking inside the ECal. For the charged-current interaction, we rely on the track multiplicity measurement. Møller and other neutrino inducing processes are hard to reject, but they are beyond the LDMX design in terms of number of electrons on target (EoT).

\section{Can we really do it?}

In order to show that we can reject backgrounds to the necessary level, we rely on detailed simulations with GEANT4~\cite{Agostinelli:2002hh,gdml}. Use simulate  photo-nuclear and conversion reactions both in the ECal and the target, labeled as "ECal PN", "Ecal conv.", "Target PN" and "Target conv.", respectively in the figures in this article.

We enhance these productions using a biasing factor that still ensures that the the longitudinal position of the decay vertex is unchanged. We use MadGraph~\cite{alwall:2007st} to simulate the dark bremsstrahlung processes with masses covering 4 orders of magnitude, and then pass the resulting kinematical information to a dedicated package with GEANT4~\cite{g4darkbrem}.

Table~\ref{tab:cutflow_with_tracking} shows the number of events left after applying selections in a cumulative way. The first four columns in the ``All / Acceptance'' row mean the number of simulated events for the backgrounds, and the last four columns show the percentage of the signal events after requiring that they are within the acceptance of the experiment. The trigger requires that the sum of the energy reconstructed in the first 20 layers of the ECal is below 3~GeV. From the triggered events, we select the ones that have beam-energy compatible momentum in the tagger tracker, requiring the incoming electron to have $p_\text{tagger} > 5.6$~GeV. The next selection requires a single track produced, as measured in the recoil tracker. 

\begin{figure}[h!]
    \centering
    \includegraphics[width=.5\textwidth]{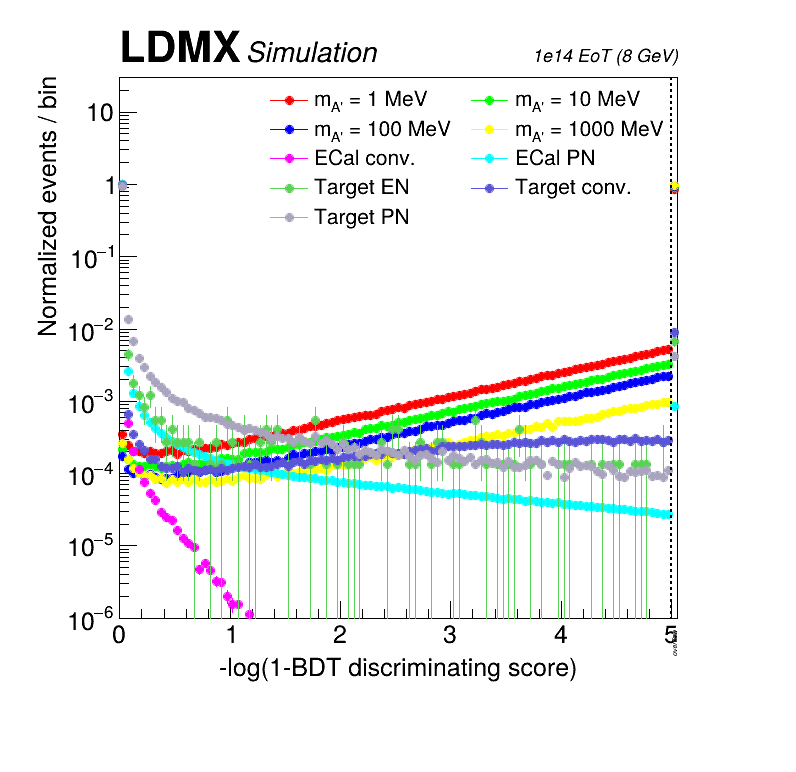}
    \caption{The logarithmic transform of the BDT scores for signal and background MC events passing the trigger and fiducial selections. The signals with different masses are displayed with red, green, blue and yellow. Magenta and cyan show the ECal conversion, and photo-nuclear samples, respectively, while the dark blue and gray show the same processes occurring in the target. Dark green is for the target electro-nuclear processes. 
    }
    \label{fig:BDTscore}
\end{figure}

We use a boosted decision tree (BDT)~\cite{10.1145/2939672.2939785} containing ECal shower and hit features to remove the complicated backgrounds in the ECal. The output of the BDT is an arbitrary score that shows how signal like an event is. The log tranform of the score is shown in Fig.~\ref{fig:BDTscore}. The selection was chosen so that we have only a handful remaining background events. In order to remove MIP tracks, like the kaons, we run a tracking algorithm inside the ECal along the projected photon path and count the number of tracks. Requiring less than 3 increases the sensitivity. In the HCal, we require that the maximum number of photo-electrons less than 8. After all these, we get about 2 background events while maintaining 50--70\% signal efficiency.

\begin{table*}[h!]
\tbl{Cutflow using the BDT approach. The first four columns display the number of simulated events for the backgrounds, and the last four columns show the percentage of the signal events passing criteria as labeled in the left.\label{tab:cutflow_with_tracking}}{
\begin{tabular}{|c||c|c|c|c||c|c|c|c|} \hline
\multicolumn{1}{|c||}{Selection}  & \multicolumn{4}{c||}{Number of Background Remaining} & \multicolumn{4}{c|}{Signal Efficiency for different $m_{A'}$} \\ \hline
& target PN  & target conv. & ECal PN & ECal conv. & 1~MeV & 10~MeV  & 100~MeV & 1000~MeV  \\ \hline 
EoT equivalent  & $10^{14}$ &  $10^{15}$  & $10^{14}$  & $10^{14}$  & & & &\\ \hline
All / Acceptance  & 1,941,296& 7,091,907 & 299,299,867& 18,720,663& 100\% & 100\% & 100\% & 100\% \\ \hline
Triggered & 756,661& 4,280,536 &   170,993,364&  14,042,193&  78.2\% & 89.6\% & 91.3\% & 94.3\% \\ \hline
$p_{\textrm{tagger}} > 5.6$~GeV & 745,494& 4,224,191 &   168,725,637&  13,856,953&  77.1\% & 88.4\% & 90.0\% &93.0\% \\ \hline
$N_{\textrm{recoil}} = 1$ & 72,927& 21,625 &   154,191,349&  12,685,692&  72.2\% & 82.4\% & 83.0\% &72.4\% \\\hline
Ecal BDT $>$ 0.9974 & 1,529& $<$1 &   64,353&  $<$1&  52.9\% &  68.6\% & 73.1\% & 67.8\% \\ \hline
$N_{straight} < 3$ & 1,412& $<$1 &   61,771&  $<$1&  51.8\% & 67.2\% & 71.7\% & 64.3\% \\ \hline
HCal maxPE $<$ 8 & $<$1   & $<$1   &  2&   $<$1& 49.9\% &  64.3\% & 68.3\% & 59.3\%\\ \hline
\end{tabular}}
\end{table*}

\section{Can we do better?}

Another approach we studied, is to rely on graph neural networks using ECal hit information, called ParticleNet~\cite{Qu:2019gqs}. All the other selections are the same as in the BDT approach. Table~\ref{tab:pnet_cutflow_table} shows the cutflow using this approach. Preselection required less than 90 hits in the ECal and the total isolated energy to be less than 1100~MeV.  In this case, the background yields are about the same as in the BDT, while the signal efficiencies are 75--95\%.

\begin{table*}[h!]
\centering
\tbl{Cutflow with the graph neural network approach. The first four columns display the number of simulated events for the back-
grounds, and the last four columns show the percentage of the signal events passing criteria as labeled in the left.\label{tab:pnet_cutflow_table}}{
\begin{tabular}{|c||c|c||c|c|c|c|} \hline
\multicolumn{1}{|c||}{Selection}  & \multicolumn{2}{c||}{Number of Background Remaining} & \multicolumn{4}{c|}{Signal Efficiency for different $m_{A'}$} \\ \hline
& eval sample (PN)  & at $10^{14}$ EoT & 1~MeV & 10~MeV & 100~MeV & 1000~MeV  \\ \hline 
Preselection & 8,929,094 & & 99.5\% & 99.8\% & 99.8\% & 99.8\% \\ \hline
ParticleNet $>$ 0.74 & 74,805 & & 92.2\% & 95.9\% & 96.4\% & 97.1\%\\ \hline
HCal maxPE $<$ 8 & 1 & 1.1 $\pm$ 0.3  & 87.6\% & 89.7\% & 89.4\% & 75.7\% \\ \hline 
\end{tabular}}
\end{table*}

 \begin{figure}[h!]
    \centering
    \includegraphics[width=.45\textwidth]{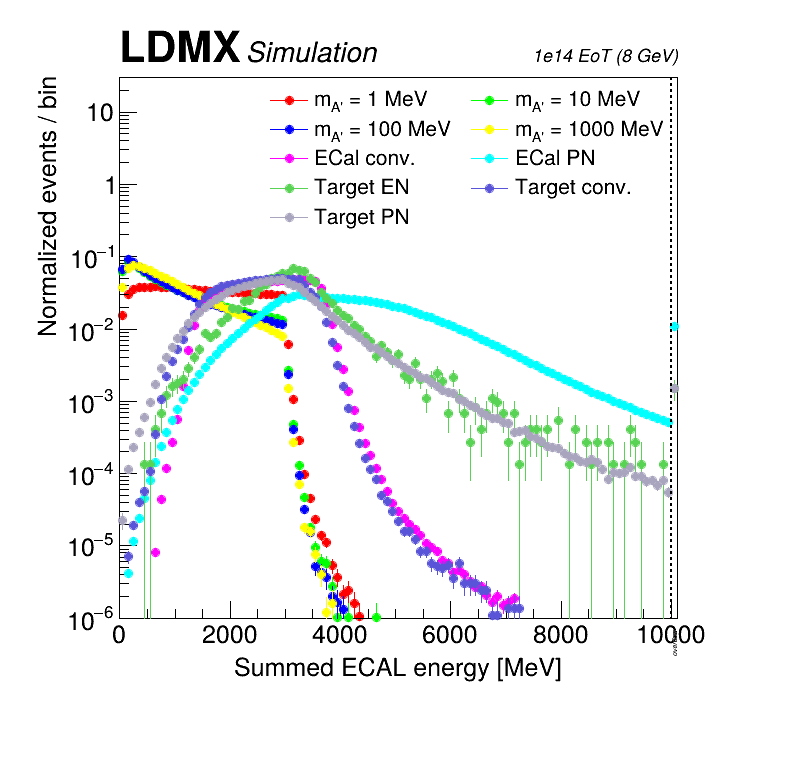}
    \caption{The summed energy deposited in the ECal after trigger selection. The signals with different masses are displayed with red, green, blue and yellow. Magenta and cyan show the ECal conversion, and photo-nuclear samples, respectively, while the dark blue and gray show the same processes occurring in the target. Dark green is for the target electro-nuclear processes.}
    \label{fig:sumEcalEnergy_readoutHits}
\end{figure}

We can also take the most important features in the BDT using the SHAP algorithm~\cite{lundberg2019consistentindividualizedfeatureattribution} and run a cut and count analysis! The variables used rely on the signals tending to have shorter, more isolated showers,
leading to less hits and activity in the ECal. As an example Fig~\ref{fig:sumEcalEnergy_readoutHits} shows the summed energy in the ECal.
In this plot, we observe the effect of the trigger, which produces a sharp drop at 3~GeV for the signal and a cut-off at 4~GeV. Background processes occur deeper in the ECal, beyond the first 20 layers used for triggering, and therefore do not exhibit a clear cut-off as the signal does. Conversion samples deposit more energy than a single recoil electron in the signal events, while the electro-nuclear (EN) and photo-nuclear events show even higher-energy tails.

We optimized the selection value for these variables using the Punzi figure-of-merit~\cite{Punzi:2003bu}. Table~\ref{tab:cncCutflowBkg} shows the cutflow using this approach. This approach still achieves $<$10 background events with a
comparable signal efficiency to the BDT.

\begin{table*}[h!]
\tbl{Cutflow with trigger, a cut and count analysis in the ECal and the HCal veto. The first four columns display the number of simulated events for the back-
grounds, and the last four columns show the percentage of the signal events passing criteria as labeled in the left.\label{tab:cncCutflowBkg}}{
\centering
\scalebox{0.95}{
\begin{tabular}{|c||c|c||c|c|c|c|} \hline
\multicolumn{1}{|c||}{Selection}  & \multicolumn{2}{|c||}{Number of Background Remaining} & \multicolumn{4}{c|}{Signal Efficiency for different $m_{A'}$} \\ \hline
& ECal PN  & ECal conversion & 1~MeV & 10~MeV & 100~MeV & 1000~MeV  \\ \hline 
EoT equivalent  & $10^{14}$ &  $10^{14}$ & & & &  \\ \hline
All / Acceptance  & 299,299,867 & 18,720,663 & 100\% & 100\% & 100\% & 100\% \\ \hline
Triggered & 170,993,364 & 14,042,193& 78.2\% & 89.6\% & 91.3\% & 94.3\%\\ \hline
$E_{\textrm{sum}} < 3500$ MeV& 56,037,910 & 12,258,095& 78.2\% & 89.6\% & 91.3\% & 94.3\%\\ \hline
$E_{\textrm{sumTight}} < 800$ MeV& 2,488,172 & 1,123,523&  73.8\% & 86.6\% & 88.3\% & 91.5\%\\ \hline
$E_{\textrm{back}} < 250$ MeV & 1,851,644 & 296&  73.7\% & 86.5\% & 88.2\% & 91.5\%\\ \hline
$N_{\textrm{hits}} < 70$ & 1,472,425 & 116& 71.6\% & 85.3\% & 87.2\% & 90.8\%\\ \hline
$RMS_{\textrm{shower}} < 110$ mm & 1,466,579 & 116& 71.5\% & 85.1\% & 86.9\% & 90.3\%\\ \hline
$E_{\textrm{cell,max}} < 300$ MeV& 1,194,631 & 114& 65.7\% & 81.0\% & 83.0\% & 85.5\%\\ \hline
$\textrm{RMS}_{\textrm{Layer,hit}} < 5$ & 1,001,030 & $<$1 & 64.4\% & 79.0\% & 81.0\% & 83.6\%\\ \hline
$N_{\textrm{straight}} < 3$ &  933,886 & $<$1 & 62.1\% & 76.2\% & 79.2\% & 73.2\%\\ \hline
HCal maxPE $<$ 8 & 7 & $<$1 & 59.4\% & 72.6\% & 75.0\% & 65.0\%\\ \hline
\end{tabular}}
}
\end{table*}

\section{Expected results}

\begin{figure}[h!]
    \centering
    \includegraphics[width=.5\textwidth]{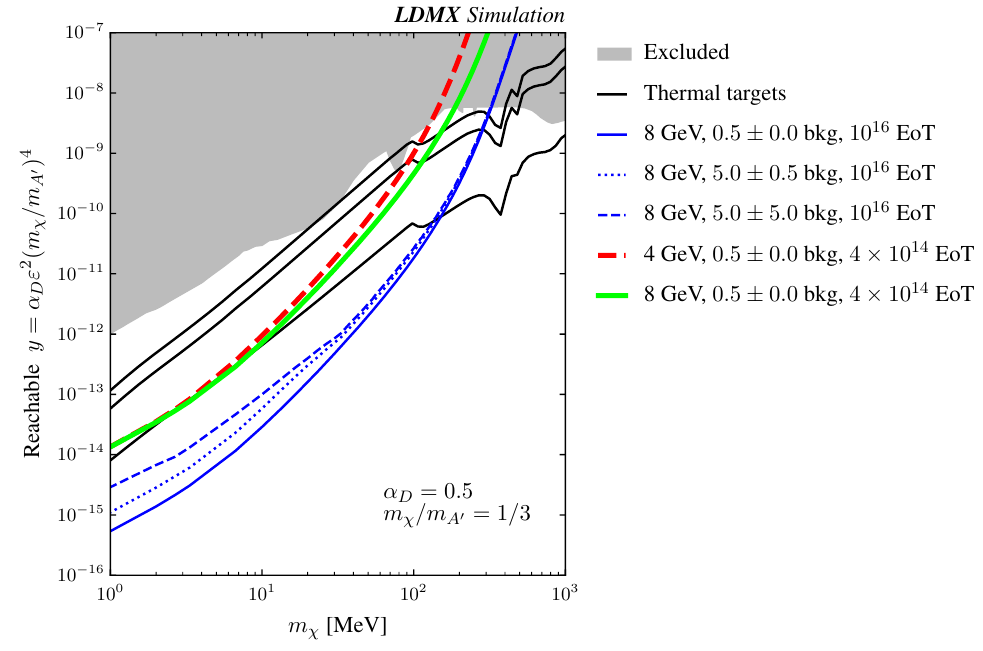}
    \caption{Expected limits of the missing-momentum search assuming a range of background events left.}
    \label{fig:reachUncert}
\end{figure}

If we do not see any signal, we can set limits on the kinematic mixing parameter ($\epsilon$) and the DM mass. We use the benchmark model where DM mass / A’ mass = 1/3. Figure~\ref{fig:reachUncert} shows the reach assuming a range of background events left denoted with different colors. The solid black lines show the thermal targets, and we can see that already at 4E14 EoT, which is about a year of runtime, we can cover a big range of them.

What if we do see signal? Given that none of the selection earlier relies on the $p_T$ of the recoil electron, we can use that information and characterize the signal~\cite{Blinov:2020epi}. We can interpret  the signal efficiency in different models, assuming a scalar, Majorana, or a pseudo-Dirac particle behavior.
Using the $p_T$ information we can determine the A' mass. 

\begin{figure}[h!]
    \centering
    \includegraphics[width=0.5\textwidth]{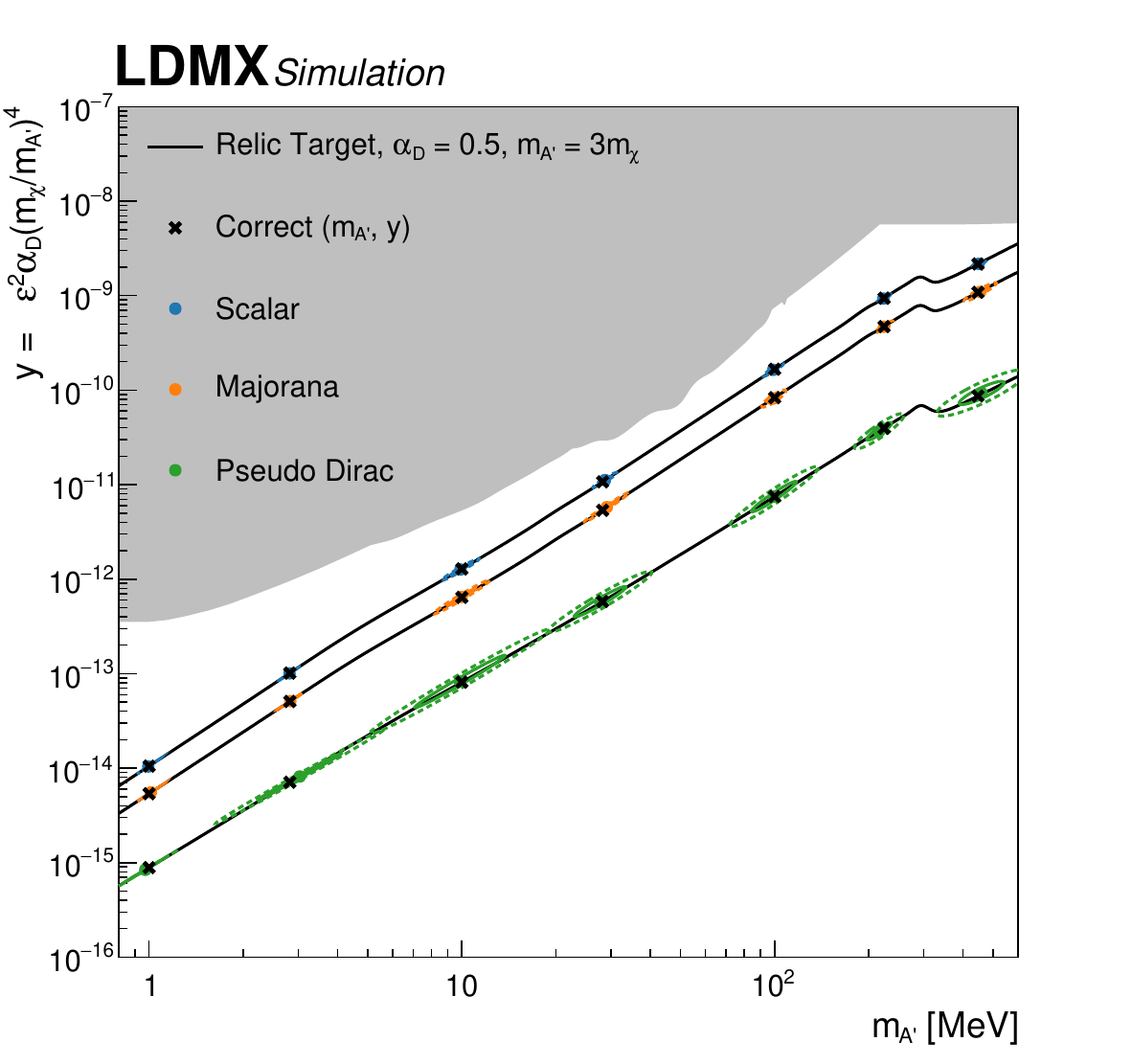}
    \caption{Comparison of simulated and reconstructed masses and couplings, including the mean and the $1\sigma$ (solid ellipses) and $2\sigma$ (dashed ellipses) confidence ellipses for different model assumptions at $1\times 10^{16}$.}
    \label{fig:model_comp}
\end{figure}

We model the signal $p_T$ spectrum using a Gaussian Process Regression (GPR) trained on simulated samples spanning $m_{A'} = 0.1~\text{MeV}–1000~\text{MeV}$, incorporating systematic effects such as $p_T$ smearing and validated against unseen mass points to ensure robustness. Using this model, we reconstruct the dark photon mass and coupling strength via a likelihood-based statistical fit~\cite{Pheno}, achieving precise mass reconstruction across a wide range of signal and background conditions. Figure~\ref{fig:model_comp} shows the simulated mass with black "x" markers, and the reconstructed mass with colors, including the 1~$\sigma$ and 2~$\sigma$ confidence ellipses. The method reconstructs masses with very good precision, and at 1E16 EoT allows us to distinguish scalar, Majorana, or pseudo-Dirac characteristics. 

\section{Other physics cases}

An alternative dark matter search channel is considered, when the ECal is used as a target. This is explored in Ref.~\cite{LDMX:2025ixw}. That work demonstrates that this analysis channel can probe previously unexplored regions of phase space even during the initial LDMX data-taking period.

Beyond the dark matter search, LDMX is able to search for axion-like particles, extend limits for milli-charged fermions and a variety of long lived models, where the DM decays in the ECal or HCal detectors~\cite{Berlin_2019}.

Furthermore, the electro-nuclear backgrounds for LDMX are very important for neutrino physics, so we developed electro-nuclear program to reduce electron-neutrino
modeling uncertainties by measuring electro-nuclear scattering precisely.

\section{Summary}

The Light Dark Matter eXperiment (LDMX) is a fixed-target experiment proposed at SLAC, designed to probe sub-GeV dark matter via missing-momentum and missing-energy signatures. Using a high-resolution tracker, a thin tungsten target, and hermetic calorimeters, LDMX can suppress Standard Model backgrounds by over 14 orders of magnitude while maintaining high signal efficiency. Detailed simulations with advanced machine learning demonstrate sensitivity to thermal dark matter models and allow characterization of potential signals. Beyond dark matter, LDMX can explore axion-like particles, milli-charged particles, long-lived particles, and precision electron–nuclear interactions, offering a versatile platform for physics beyond the Standard Model.

\bibliography{PHEP_LDMX}

\end{document}